\documentclass[12pt,indent]{article}%a5paper
\usepackage{a4,latexsym,amsmath}%a4
\usepackage[latin1]{inputenc}
\usepackage{float}
\usepackage{natbib}
\usepackage{graphics}
\usepackage[english]{babel}
\usepackage{tabularx}
\usepackage{lscape}
\usepackage{multirow}
\usepackage{rotating}
\usepackage{dsfont}
\usepackage{subfig}
\usepackage{bbm}
\usepackage[table]{xcolor}

\usepackage{amsmath}
\usepackage{amsfonts}

\usepackage{calc}
\usepackage{ifthen}

\usepackage{%
  longtable,
  array,
  booktabs,
  dcolumn,
  rotating,
  shortvrb,
  tabularx,
  units,
  multirow
}

\newcommand{\alphab}{\boldsymbol{\alpha}}

\newcommand{\betab}{{\boldsymbol{\beta}}}

\newcommand{\gammab}{\boldsymbol{\gamma}}

\newcommand{\xb}{\boldsymbol{x}}

\newcommand{\zb}{\boldsymbol{z}}

\newcommand{\blanco}[1]{}

\def\d{\displaystyle}

\usepackage{natbib}

\usepackage{setspace}
\begin{document}
\bibliographystyle{chicago}
\sloppy
%%%%%%%%%%%%%%%%%%%%%%%%%%%%%%%%%%%%%%%%%%%%%%%%%%%%%%%%%%%%%%%%%%%%%
%                                                                   %
%         Definition einer modifizierten Kapitelï¿½berschrift         %
%                                                                   %

\makeatletter
\renewcommand{\section}{\@startsection{section}{1}{\z@}%
        {-3.5ex \@plus -1ex \@minus -.2ex}%
        {1.5ex \@plus.2ex}%
        {\reset@font\Large\sffamily}}
\renewcommand{\subsection}{\@startsection{subsection}{1}{\z@}%
        {-3.25ex \@plus -1ex \@minus -.2ex}%
        {1.1ex \@plus.2ex}%
        {\reset@font\large\sffamily\flushleft}}
\renewcommand{\subsubsection}{\@startsection{subsubsection}{1}{\z@}%
        {-3.25ex \@plus -1ex \@minus -.2ex}%
        {1.1ex \@plus.2ex}%
        {\reset@font\normalsize\sffamily\flushleft}}
\makeatother

%                                                                   %
%%%%%%%%%%%%%%%%%%%%%%%%%%%%%%%%%%%%%%%%%%%%%%%%%%%%%%%%%%%%%%%%%%%%%

%%%%%%%%%%%%%%%%%%%%%%%%%%%%%%%%%%%%%%%%%%%%%%%%%%%%%%%%%%%%%%%%%%%%%
%                                                                   %
%         Definition einer modifizierten Bildunterschrift           %
%                                                                   %

\newsavebox{\tempbox}
\newlength{\linelength}
\setlength{\linelength}{\linewidth-10mm} \makeatletter
\renewcommand{\@makecaption}[2]
{
  \renewcommand{\baselinestretch}{1.1} \normalsize\small
  \vspace{5mm}
  \sbox{\tempbox}{#1: #2}
  \ifthenelse{\lengthtest{\wd\tempbox>\linelength}}
  {\noindent\hspace*{4mm}\parbox{\linewidth-10mm}{\sc#1: \sl#2\par}}
  {\begin{center}\sc#1: \sl#2\par\end{center}}
}

%                                                                   %
%%%%%%%%%%%%%%%%%%%%%%%%%%%%%%%%%%%%%%%%%%%%%%%%%%%%%%%%%%%%%%%%%%%%%

%\bibliographystyle{chicago}
%\baselineskip7mm
%\parindent 0.5cm
%\parskip2ex plus0.5ex minus 0.5ex
%\setlength{\parskip}{7pt plus 1pt minus 1pt}

\def\R{\mathchoice{ \hbox{${\rm I}\!{\rm R}$} }
                   { \hbox{${\rm I}\!{\rm R}$} }
                   { \hbox{$ \scriptstyle  {\rm I}\!{\rm R}$} }
                   { \hbox{$ \scriptscriptstyle  {\rm I}\!{\rm R}$} }  }

\def\N{\mathchoice{ \hbox{${\rm I}\!{\rm N}$} }
                   { \hbox{${\rm I}\!{\rm N}$} }
                   { \hbox{$ \scriptstyle  {\rm I}\!{\rm N}$} }
                   { \hbox{$ \scriptscriptstyle  {\rm I}\!{\rm N}$} }  }

\def\d{\displaystyle}

\title{Tree-Structured Scale Effects in Binary and Ordinal Regression }
\author{Gerhard Tutz$^*$    \& Moritz Berger$^{**}$ \vspace{0.4cm}\\
{\small $^*$Ludwig-Maximilians-Universit\"{a}t M\"{u}nchen}\\
{\small $^{**}$ Institut für Medizinische Biometrie,}\\ {\small Informatik und Epidemiologie, Universitätsklinikum Bonn}}

%\author{Gerhard Tutz \\{\small Ludwig-Maximilians-Universit\"{a}t M\"{u}nchen}\\
%{\small Akademiestra{\ss}e 1, 80799 M\"{u}nchen}}

%\author{Jan Gertheiss\footnote{To whom correspondence should be
%addressed: \texttt{jan.gertheiss@stat.uni-muenchen.de.}}
%\footnote{Department of Statistics, Ludwig-Maximilians-Universität
%Munich, Germany.} \ \& Gerhard Tutz\footnotemark[2]}

% \ead{tutz@stat.uni-muenchen.de}
% \address{Ludwig-Maximilians-Universit\"{a}t M\"{u}nchen, Ludwigstra{\ss}e 33, D-80539 M\"{u}nchen, Germany}
%\author{Lorenz Uhlmann}
%\ead{tutz@stat.uni-muenchen.de}
%\address[muc]{Ludwig-Maximilians-Universit\"{a}t M\"{u}nchen, Akademiestra{\ss}e 1, 80799 M\"{u}nchen, Germany}
%\cortext[cor]{Corresponding author. Tel.: ++4989 2180 3044; fax.:
%++4989 2180 5308.}
%{\texttt{\small \{tutz, uhlmann\}@stat.uni-muenchen.de}}}
%\address[muc1]{Ludwig-Maximilians-University Munich, Ludwigstrasse 33, D-80539 Munich, Germany}
%\address[muc2]{Ludwig-Maximilians-University Munich, Akademiestra{\ss}e 1, D-80799 Munich, Germany}
%\cortext[cor]{Corresponding author. Tel.: ++49 89 2180 3044; fax.:
%++49 89 2180 ???.}
\maketitle

\begin{center}\small
%File: PropOddstree (TuHeterogeneousTree)
\end{center}

\begin{abstract} % \renewcommand{\baselinestretch}{1.3} \small\normalsize
\noindent
%\begin{center}
In binary and ordinal regression one can distinguish between a location component and a scaling component. While the former determines the location within the range of the response categories, the scaling indicates  variance heterogeneity. In particular since it has been demonstrated that misleading effects can occur if one ignores the presence of a scaling component it is important to account for potential scaling effects in the regression model, which is not possible in available recursive partitioning methods. The proposed recursive partitioning method yields two trees, one for the location and one for the scaling. They show in a simple interpretable way how variables interact to determine the binary or ordinal response. The developed algorithm controls for the global significance level and automatically selects the variables that have an impact on the response. The modelling approach is illustrated by several real-world applications.

\end{abstract}


\section{Introduction}
Tree-based models are strong non parametric tools that allow to investigate interaction effects of covariates on responses.
The basic concept is very simple: By binary recursive partitioning the predictor space is partitioned into a set of rectangles and on each rectangle a simple model (for
example a constant) is fitted. The most popular versions are CART \citep{BreiFrieOls:84}, which is an abbreviation for classification and regression trees, and conditional inference trees, abbreviated by CTREE \citep{Hotetal:2006}. Introductions and overviews were given, among others, by \citet{loh2014fifty} and \citet{Strobetal:2009}.
Recursive partitioning methods, or simply trees, have several advantages: (i) they can be used in high dimensional settings  because they provide automatic variable selection, (ii) they have a built-in interaction detector, and (iii) they are easy to interpret and visualize. Besides classical regression trees for metrically scaled response variables, also versions for binary and ordinal responses are available, see \citet{Piccarreta:2008}, \citet{archer2010rpartordinal} and \citet{galimberti2012classification}. 

The objective of the present paper is to introduce trees in regression structures with ordinal responses that include scale effects, which are needed if unobserved heterogeneity of variances is present. The modelling of scale effects in ordinal regression was already considered by \citet{McCullagh:80}, who introduced the so-called \textit{location-scale model} and gave a simple example with one binary covariate dealing with the quality of right eye vision for men and women. The location-scale model was  considered and extended, among others,  by \citet{cox1995location} and \citet{tutz2017separating}; \citet{ishwaran2000general} investigated the link to ROC analysis, \citet{hedeker2008application}, \citet{hedeker2009mixed} and \citet{hedeker2012modeling} showed how to use it in the case of repeated ordinal measurements.  

Scale effects are also found in binary data. Their potential impact  found much attention since \citet{allison1999comparing} demonstrated that comparisons of binary model coefficients across groups can be misleading if one has underlying heterogeneity of residual variances. The problem has been investigated in various papers since then, see \citet{williams2009using}, \citet{mood2010logistic}, \citet{karlson2012comparing}, \citet{breen2014correlations} and \citet{rohwer2015note}.
One strategy to account for heterogeneity is to use McCullagh's location-scale model, which in the social sciences is also known as the heterogeneous choice  or heteroskedastic logit model \citep{alvarez1995american, williams2009using}. It is included in various program packages as Stata, Limdep, SAS, and R. 

%In the implementation of our proposed tree-structured algorithm (see Section \ref{algorithm}) we make use of the R package \textbf{ordinal} for fitting models.   

As a parametric model that uses linear predictors the location-scale model is rather restrictive. In particular interactions of higher order are hard to include and lower order interactions are restricted to linear interactions. Tree-based methods offer a non parametric alternative to investigate the interaction structure 
and automatically  select variables. Variable selection is important since typically it is not known which variables contribute to location and to scaling. Since  there are two components in the model, location and scaling, classical recursive partitioning methods can not be used. The method developed in the following is explicitly designed to account for these two components. Two separate trees are obtained,  one for each component.

In Section \ref{sec:tree} the basic approach is introduced and illustrated by an application. In Section \ref{algorithm} the proposed algorithm is given in detail.
More applications are considered in Section \ref{sec: appl}. The paper concludes with a summary given in Section~\ref{sec:summ}.

%The social science version in the form of the  heterogeneous choice model was used in particular by . 

\section{Trees with Scale Effects}\label{sec:tree} 
In the following we first consider  basic ordinal models  and the problems that might occur if variance heterogeneity is ignored. Then we introduce the tree-structured modelling approach that is proposed. 

\subsection{Proportional Odds and Location-Scale Model}

A common way to derive ordinal regression models is to assume that a latent variable is behind the ordinal response $Y$. Let the latent regression model 
have the form
\[
Y_i^*=\alpha_0+ \xb_i^T \alphab+\sigma\varepsilon_i,\quad i=1,\hdots,n\,, %  x_{i1}\alpha_{1}+\dots+x_{ip}\alpha_p+\sigma\varepsilon_i,
\]
where $Y_i^*$ is the latent variable, $\xb_i$ is a vector of covariates, and $\sigma$ is the standard deviation of the noise variable $\varepsilon_i$, which has symmetric distribution function~$F(.)$.
The essential concept is to consider the ordinal response as a categorized version of
the latent variable with the link between the observable ordinal variable $Y_i$ with $k$ categories and the latent variable $Y_i^*$ given by
\begin{equation}\label{eq:catbound}
Y_i=r\quad\Leftrightarrow\quad\theta_{ r-1}\,<\,Y_i^*\,\leq\,\theta_{r}\,,
\end{equation}
where $-\infty =\theta_{0}<\theta_{1}<\dots<\theta_{k}=\infty$
are thresholds on the latent scale. Simple derivation yields that the response probabilities are   given by
\[
P(Y_i\leq r|\xb_i)=F\left(\frac{\alpha_{0r}-\xb_i^T\alphab}{\sigma}\right)\,,
\]
where $\alpha_{0r}=\theta_{r}-\alpha_{0}$. However, the   model   parameters are not identifiable. An identifiable version is obtained by setting $\sigma=1$ or, equivalently, using $\beta_{0r}=\alpha_{0r}/\sigma$, $\betab=\alphab/\sigma$, which yields 
the  \textit{cumulative model}
\begin{equation}\label{eq:cum}
P(Y_i\leq r|\xb_i)=F({\beta_{0r}-\xb_i^T\betab})\,.
\end{equation}
The most prominent member of the family of cumulative models is the \textit{proportional odds model}, which uses the logistic distribution function $F(\eta)=\exp(\eta)/(1+\exp(\eta))$. It has the form
\begin{equation}\label{eq:ppo}
\log\left(\frac{P(Y_i\leq r|\xb_i)}{P(Y_i > r|\xb_i)}\right)=\eta_{ir}={\beta_{0r}-\xb_i^T\betab}\,.
\end{equation}
The strength of model \eqref{eq:ppo} is that the parameters have an easily accessible interpretation. Let $\gamma_r(\xb_i)=P(Y_i > r|\xb_i)/P(Y_i \leq r|\xb_i)$ denote the cumulative odds for category $r$. Then one can derive that the effect of the $j$th variable is given by
\begin{equation}\label{eq:int}
e^{\beta_j}=\frac{\gamma_r(x_{i1},\dots,x_{ij}+1,\dots,x_{ip})}{\gamma_r(x_{i1},\dots,x_{ij},\dots,x_{ip})}\,,
\end{equation}
which does not depend on $r$. That means that $e^{\beta_j}$ represents the multiplicative change in cumulative odds if $x_{ij}$ increases by one unit for each category. Of course, the interpretation holds only if the model holds or is at least a good approximation to the data generating model.  

It has been shown that the cumulative model \eqref{eq:cum} can yield very misleading results if there is  variance heterogeneity in the underlying continuous regression model. 
\citet{allison1999comparing} considered an example with the binary response being the promotion to an associate professor from the assistant professor level. It turned out that the number of published articles had a much stronger effect for male researchers than for female researchers, which seems rather unfair. He demonstrated that this effect could be due to heterogeneous variances. 

The effect of heterogeneous variances is easily seen. Let the latent regression model be given by $Y_i^*=\alpha_0+ \xb_i^T \alphab+\sigma_i\varepsilon_i$, where $\sigma_i$ now depends on the specific observation $i$. In the simplest case one has $\sigma_i=z_i\gamma$, where $z_i$ is an indicator variable, which takes the value one for group 1 (for example males) and the value zero for group 0 (for example females). Then the simple cumulative model \eqref{eq:cum} is mis-specified. The derivation from the latent variable yields 
\begin{equation}
\begin{aligned}
&P(Y_i\leq r|\xb_i)=F({\alpha_{0r}/\sigma-\xb_i^T(\alphab}/{\sigma})) \quad \text{for observations from group 1 and}\\
&P(Y_i\leq r|\xb_i)=F({\alpha_{0r}-\xb_i^T\alphab}{}) \qquad \qquad \text{for observations from group 0}\,.
\end{aligned}
\end{equation}
Thus, effects of covariates differ between the groups. One has $\alphab/\sigma$ in group 1 and $\alphab$ in group 0. If, for example, $\sigma=0.5$ the effect strength in group 1 is twice the effect strength in group 0. The dependence on the group is simply ignored if one sets $\sigma=1$, which is typically assumed in categorical regression. It means that in both groups   the same scaling is used, although   different ones are needed, see also \citet{williams2009using}, \citet{mood2010logistic}.

This form of mis-specification can be avoided by explicit modelling of the heterogeneity of variances. Let the standard deviation be determined by $\sigma_i=\exp(\zb_i^T\gammab)$, where $\zb_i$ is an additional vector of covariates, then one obtains from assumption \eqref{eq:catbound} the \textit{location-scale model}  

\begin{equation}\label{eq:hetchoice}
P(Y_i\le r|\xb_i,\zb_i)=F\left(\frac{\beta_{0r}-\xb_i^T\betab}{\exp(\zb_i^T\gammab)}\right)\,,
\end{equation}
which for the logistic distribution function yields
\begin{equation}\label{eq:hetchoice}
\log\left(\frac{P(Y_i\le r|\xb_i,\zb_i)}{P(Y_i> r|\xb_i,\zb_i)}\right)=\eta_{ir}=\frac{\beta_{0r}-\xb_i^T\betab}{\exp(\zb_i^T\gammab)}\,.
\end{equation}
The model contains two  terms in the predictor that specifies the impact of covariates. The first is  the location term $\beta_{0r}+\xb_i^T\betab$, the second is the variance or scaling term $\exp(\zb_i^T\gammab)$, which derives from the ``variance equation'' $\sigma_i=\exp(\zb_i^T\gammab)$.
Importantly, if $\xb_i$ and $\zb_i$ are distinct the interpretation of the $\xb$-variables is the same as in the proportional odds model. 
With $\gamma_r(\xb_i,\zb_i)=P(Y_i > r|\xb_i,\zb_i)/P(Y_i \leq r|\xb_i,\zb_i)$ denoting the cumulative odds for category $r$ one obtains again the relation \eqref{eq:int} and therefore an interpretation of parameters that does not depend on the category.

The location-scale model was introduced by \citet{McCullagh:80} but is also known as \textit{heterogeneous choice model} or heteroscedastic logit model \citep{alvarez1995american}. It should be noted that although the scaling component is typically motivated from variance heterogeneity it can also be seen as   representing interactions or effect-modifying effects, see \citet{rohwer2015note}, \citet{Tu2018Het}. As \citet{williams2010fitting} noted, it is also strongly related to  
the logistic response model with proportionality constraints  proposed by \citet{hauser20061} and extended by \citet{fullerton2012proportional}.

\subsection{Tree-Structured Location-Scale Models}
Recursive partitioning methods for ordinal responses have been proposed by \citet{archer2010rpartordinal},  \citet{galimberti2012classification}, \citet{JanTuBoul16} and are available in R packages. Also the conditional unbiased recursive partitioning framework as proposed by \citet{Hotetal:2006} allows to fit trees for ordinal responses. However, all of these methods do not account for possible heterogeneity induced by variance. 

The problem with modelling heterogeneity is that one has to fit two separate predictors, the location term and the variance term. In the traditional location-scale model \eqref{eq:hetchoice} they are represented by the linear predictor $\beta_{0r}-\xb_i^T\betab$ and the variance term $\exp(\zb_i^T\gammab)$, respectively. The tree proposed here also distinguishes between location and variance; for both components separate trees are fitted. It is crucial that  the partitioning of location and variance terms has to be done in a coordinated way. Trees have to be grown by taking both components into account simultaneously.

In the following, we first sketch the basic algorithm, which will be given in more detail in Section \ref{algorithm}. The basic concept is to replace the predictor $\eta_{ir}=(\beta_{0r}-\xb_i^T\betab)/{\exp(\zb_i^T\gammab)}$ of the location-scale model \eqref{eq:hetchoice} by coordinated recursive partitioning terms. 

\subsubsection*{Basic Algorithm}

\noindent Let us consider the building of a tree when starting at the root. We will focus on metrically scaled and ordinal (including binary) covariates. In this case the partition of a node $A$ into two subsets $A_1$ and $A_2$ has the form
\[
A_1 = A \cap \{ x_j \leq c \} \quad \text{and} \quad  A_2= A \cap \{ x_j > c \}\,,
\]
with regard to threshold $c$ on variable $x_j$.

\vspace{0.2cm}
\hrule
\vspace{0.2cm}

\medskip\noindent \textit{First Step}\medskip

\noindent For each variable $x_j$ and all corresponding thresholds $c$ that can be built for this variable one investigates the following fits: 

\begin{enumerate}
\item[(a)] Location term:

One fits the location-scale model with one split in the location term and predictor 
\[
\eta_{ir} =  {\beta_{0r}- \beta  I(x_{ij} \le c)}\,, 
\]
where $I(.)$ is the indicator function. Then one obtains
\begin{align*} 
&\eta_{ir} =   {\beta_{0r}- \beta  }  \quad   \text{if} \; x_{ij} \le c \quad \text{and}\\
&\eta_{ir} =   \beta_{0r}    \qquad \;\;\; \text{if} \; x_{ij} > c \,.
\end{align*}

Alternatively one can replace $I(.)$ by $I^*(.)=2I(.)-1$, which means one uses effect coding and replaces the 0$-$1 dummy variable by the variable $I^*(.)=1$ if $x_{ij} \le c$ and $I^*(.)=-1$ otherwise. Accordingly one obtains
\begin{align*} 
\eta_{ir} &=   {\beta_{0r}- \beta  }   \quad \text{if} \; x_{ij} \le c \quad \text{and}\\
\eta_{ir} &=   \beta_{0r}+ \beta    \quad  \text{if} \; x_{ij} > c\,.
\end{align*}

\item[(b)] Variance Term:

One fits the location-scale model with one split in the variance term and predictor 
\[
\eta_{ir} = \frac{\beta_{0r}}{{\exp(\gamma  I(x_{ij} \le c)})}. 
\]
Then one obtains
\begin{align*} 
&\eta_{ir} =   \frac{\beta_{0r}}{\exp(\gamma)}  \quad   \text{if} \; x_{ij} \le c \quad \text{and}\\
&\eta_{ir} =   \beta_{0r}    \qquad \;\;\; \text{if} \; x_{ij} > c \,.
\end{align*}

\end{enumerate}

\noindent One chooses the best split according to an appropriate splitting criterion (for details, see Section \ref{algorithm}) among all the fitted models from (a) and (b). Thus in the first step one split is performed either in the location term or the variance term.

\medskip
\noindent \textit{Later Steps} \medskip

\noindent In later steps the splitting is done in a similar way. Let  $A_1^{\text{loc}},\dots,A_{m_\text{loc}}^{\text{loc}}$ denote the nodes (subsets of the predictor space) of the location term from the previous steps. Accordingly, let  $A_1^{\text{sc}},\dots,A_{m_\text{sc}}^{\text{sc}}$ denote  the nodes (subsets of the predictor space) of the variance term from the previous steps. 
Note, that all nodes are determined by a product of indicator functions. For example, if the splits were in the metric variables $x_3$ and $x_7$ a node may be determined by $I(\xb_i \in A)=I(x_{i3} > 20)I(x_{i7} \le 4)$. 

\noindent One fits all the candidate models 

\begin{enumerate}
\item[(a)] for the splitting of $A_{k}^{\text{loc}},\,k=1,\hdots,m_{loc}$ in the location term with predictors
\[
\eta_{ir} =  \frac{\beta_{0r} - \sum_{s=1}^{m_{loc}} \beta_s  I(\xb_i \in A_s^{\text{loc}}) - \beta  I(\xb_i \in A_k^{\text{loc}})I(x_{ij} \le c)}{\exp\left(\sum_{\ell=1}^{m_\text{sc}}\gamma_{\ell} I(\xb_i \in A_{\ell}^{\text{sc}})\right) }\,   
\]
to obtain  the $(m_{loc}+1)$-th node in the location term with parameter estimate~$\beta$, 

\item[(b)] for the splitting of $A_k^{\text{sc}},\,k=1,\hdots,m_{sc}$ in the variance term with predictor
\[                                                           
\eta_{ir} =  \frac{\beta_{0r}- \sum_{s=1}^{m_\text{loc}}\beta_s I(\xb_i \in A_s^{\text{loc}})}{\exp(\sum_{\ell=1}^{m_\text{sc}}\gamma_{\ell} I(\xb_i \in A_{\ell}^{\text{sc}}) + \gamma I(\xb_i \in A_k^{\text{sc}})I(x_{ij} \le c))}\,.
\]
to obtain  the $(m_{sc}+1)$-th node in the variance term with parameter estimate~$\gamma$. 
\end{enumerate}
\noindent One chooses the best split according to an appropriate splitting criterion %(for details, see Section \ref{algorithm}) 
among all the possible models from (a) and (b). Again, each step means an update of the location term or the variance term. 
After termination of the algorithm according to an appropriate stopping criterion, the final model consists of two trees, one for the location component and one for the scale component, with different partitions. 

\vspace{0.2cm}
\hrule
\vspace{0.6cm}

\noindent We   refer to the concept as \textit{tree-structured model building} to distinguish it from the \textit{model-based}  recursive partitioning models as considered by \citet{zeileis2008model}. The basic idea of model-based recursive partitioning is to fit models  in subspaces of the predictor space and then decide which partitioning explains the predictor-response relationships best. Of course elaborated methods are needed to ensure that the splits represent relevant information, for example, by using appropriate tests,  see \citet{zeileis2008model}. Although  in principle this approach could  also be used  in the location-scale framework the obtained tree would not separate between the the two types of influential terms.
The main difference between tree-structured modelling and model-based recursive partitioning  is that  tree-structured model building means that \textit{the predictor structure is determined by trees}, whereas model-based approaches do not structure the predictor but fit the whole model in sub spaces.  Model-based trees yield separate trees for the two influential terms, one tree for the location and one tree for the variance heterogeneity. Thus, it is easily seen which variables contribute to  which component. Tree structures in the predictor have been considered before, but in a quite different context;   %\citet{BergTu2016DIF}
\citet{berger2017tree} and \citet{tutz2018tree} considered trees to model  the effect of categorical predictors on the response  if the predictors have a very large number of categories.

Before considering an illustrative example we briefly consider the interpretation of parameters.
Let $A_1^{\text{loc}},\dots,A_{m_{loc}}^{\text{loc}}$ denote the end nodes of the location term, and $A_1^{\text{sc}},\dots,A_{m_{sc}}^{\text{sc}}$ denote the end
nodes of the variance term. Then one has the predictor  
\[                                                           
\eta_{ir} =  \frac{\beta_{0r}- \sum_{s=1}^{m_{loc}}\beta_s I(\xb_i \in A_s^{\text{loc}})}{\exp(\sum_{\ell=1}^{m_{sc}}\gamma_{\ell}  I(\xb_i \in A_{\ell}^{\text{sc}}))}, r=1,\dots,k-1\,. 
\]
The interpretation is similar to the interpretation of parameters in the location-scale model, the $\beta$-parameters indicate the location  and the 
$\gamma$-parameters variance heterogeneity. For illustration let us consider extreme cases. 

\begin{itemize}
\item[--]
If $\beta_s \rightarrow -\infty $ one obtains for $\xb_i \in A_s^{\text{loc}}$ (fixed variance component) the probabilities $P(Y_i=1|\xb_i)=1$, and $P(Y_i=2|\xb_i)=
\dots P(Y_i=k|\xb_i)=0$. If $\beta_s \rightarrow  \infty $ one obtains for $\xb_i \in A_s^{\text{loc}}$ the probabilities $P(Y_i=k|\xb_i)=1$, and $P(Y_i=1|\xb_i)=
\dots P(Y_i=k-1|\xb_i)=0$. That means the size of $\beta_s$ indicates the preference for high categories.

\item[--]
If $\gamma_{\ell} \rightarrow \infty $ one obtains for $\xb_i \in A_{\ell}^{\text{loc}}$ (fixed location component) the probabilities $P(Y_i=1|\xb_i)= P(Y_i=k|\xb_i)=0.5$, that means maximal heterogeneity with all responses in the extreme categories. 
\end{itemize}

\subsection{Illustrative Example}

\subsubsection*{Confidence Data}
We consider data from the general social survey of social science, in short ALLBUS, a study by the German institute GESIS. The data is available from \texttt{http://www.gesis.org/allbus}. Our analysis is based on a subset containing 2935 respondents of the ALLBUS in 2012. The response is the confidence in the federal government measured on a symmetric scale from 1 (no confidence at all/excessive distrust) to 7 (excessive confidence). As explanatory variables we consider the gender (0: male, 1: female), the income in thousands of Euros, the age in decades (centered at 50) and the self reported interest in politics from 1 (very strong interest) to 5 (no interest at all). 
%For modelling we chose category \grqq no\grqq{} as reference. The deviance of the location-shift model (without scaling) is $10,179.51$. For the model with scaled shifting of thresholds one obtains a remarkably smaller value of $10,140.91$. Hence we will present results for the model with scaling. The likelihood ratio test statistic for the null hypotheses $H_0:\alphab=\0$ is $54.5$ on 8 degrees of freedom and therefore dispersion should definitely be taken into account.
\begin{figure}[!t]%[!h]
	\centering
		\includegraphics[width=1\textwidth, trim= 3cm 2cm 3cm 0cm]{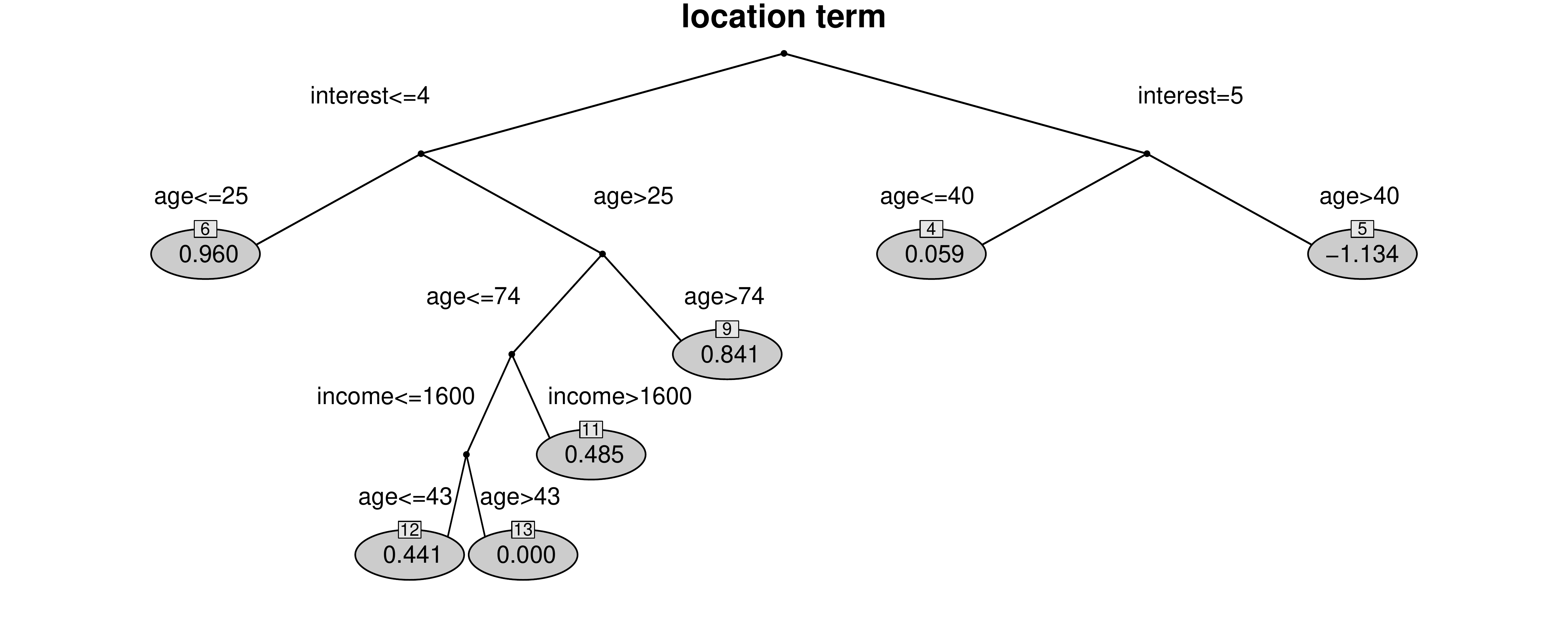}
	\caption{Tree for location term of confidence data. The parameter estimates~$\hat{\beta}_s$ are given in the terminal nodes.}
  \label{fig:confbeta}
\end{figure}

\begin{figure}[!t]%[!h]
	\centering
		\includegraphics[width=1\textwidth, trim= 2cm 2cm 2cm 0cm]{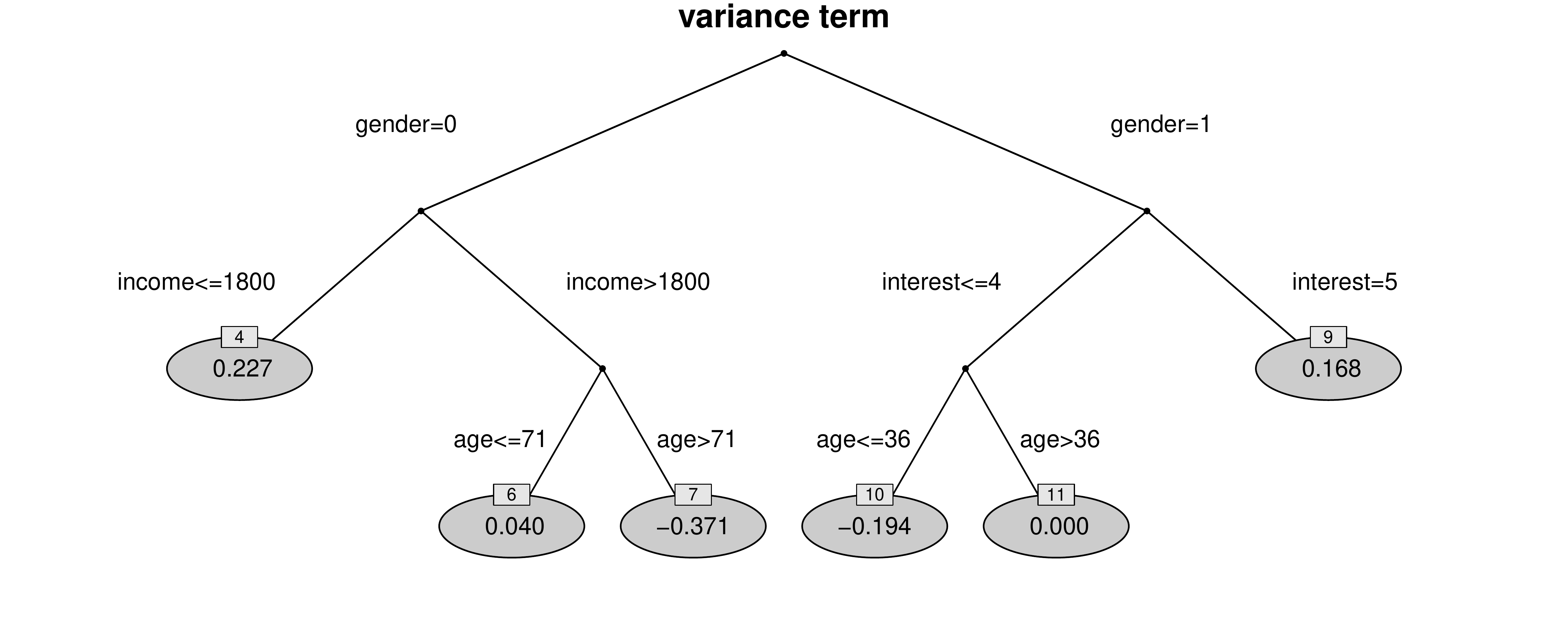}
	\caption{Tree for variance term of confidence data. The parameter estimates~$\hat{\gamma}_\ell$ are given in the terminal nodes.}
  \label{fig:confalpha}
\end{figure}

Figure \ref{fig:confbeta} shows the tree obtained for the location term and Figure \ref{fig:confalpha} the tree for the variance term. It is seen that the main drivers of confidence are interest in politics and age. Among respondents that have strong interest in political issues (interest=5) those above 40 years of age have weak confidence (node 5) whereas those below 40 years tend to prefer higher categories (node 4). Among respondents that are less interested in politics in particular young people (age lesser than 25) and older people (age above 74) show a strong tendency to choose high confidence categories ($\hat{\beta}_s=0.960$ and $\hat{\beta}_s=0.841$).
From the variance tree it is seen  that females with high interest (node 9; $\hat{\gamma}_\ell=0.168$) and males with low income (node 4; $\hat{\gamma}_\ell=0.227$) are the most heterogeneous groups with comparatively large variance. Older males with high income (node 7) followed by younger females with moderate interest form the most homogeneous groups.

\section{The Algorithm in Detail} \label{algorithm}

In all tree-based methods, one has to decide in particular how to split and how to determine
the size of the trees. In traditional approaches, one typically grows large trees and prunes them to an adequate size afterwards, see \citet{BreiFrieOls:84} and \citet{Ripley:96}. An alternative strategy, which was propagated within the conditional unbiased recursive partitioning framework  \citep{Hotetal:2006},
is to directly control the size of the trees by early stopping. We also use this approach and control the significance of splits by using tests for cumulative regression models. % it is quite natural to use test-based splits. 

Let us consider again the construction of the first split. A split in the location term with regard to the $j$-th variable yields the model with predictor 
\[
\eta_{ir} =  {\beta_{0r} - \beta_j  I(x_{ij} \le c_j)}\,, 
\]
a split in the variance term with regard to the $j$-the variable yields the model with predictor
\[
\eta_{ir} = \frac{\beta_{0r}}{{\exp(\gamma_j  I(x_{ij} \le c_j)})}\,. 
\]
To test for the best split among all the covariates, the set of possible split points and the two components (location or variance) one examines all the null hypotheses $H_0: \beta_j=0$ and $H_0: \gamma_j=0$ and selects that split as the optimal one that has the smallest $p$-value. As test statistic, we use the LR test statistic.
Computing the LR test statistic requires fitting of  both models, the full
model and the restricted model under $H_0$. We nevertheless prefer the LR statistic
because it corresponds to selecting the model  with minimal deviance. This criterion
 is also equivalent to minimizing the entropy, which belongs
to the family of impurity measures.

To decide whether the selected split should be performed, we apply a concept based on maximally selected statistics. The basic idea is to investigate the dependence of the
ordinal response and the selected variable at a global level that takes the number of splits   into account. For one fixed component and variable $j$,  one simultaneously considers all LR test statistics $T_{jc_j}$, where $c_j$ are from the set of possible split points, and computes the maximal value statistic $T_j=\max_{c_j}T_{jc_j}$. The $p$-value that can be obtained by the distribution of $T_j$ provides a measure for the relevance of variable $j$. The result is not influenced by the number of split points  since they have been   taken into account  yielding unbiased selection; for similar approaches, which inspired the proposed method, see \citet{HotLau:03}, \citet{Shih:04}, \citet{ShiTsa:2004}, \citet{StrBouAug:2007}. The method explicitly accounts for the involved multiple testing problem. As the distribution of $T_j$ in general is unknown we use a permutation test to obtain a decision on the null hypothesis. The distribution of $T_j$ is determined by computing the maximal value statistics based on random permutations of variable $j$. A random permutation of variable $j$ breaks the relation of the covariate and the response in the original data. By computing the maximal value statistics for a large number of permutations one obtains an approximation of the distribution under the null hypothesis and the corresponding $p$-value. Importantly, to determine the $p$-value with sufficient accuracy, the number of permutations should increase with the number of covariates.

In all later steps the basic procedure is the same, one searches for the
statistic with the maximal value trying all combinations of variables and split points in both components. For the components that already have been split (location, variance or both) one starts from already built nodes. 
Given overall significance level $\alpha$  the significance level for the permutation test that tests splits in one variable is chosen by $\alpha/p$, where $p$ denotes the number of covariates that are available. 

Alltogether, the following steps are carried out during the fitting procedure:

\begin{enumerate}
\item {\it (Initial Model)}. Fit the model with category-specific intercepts only, yielding the estimates $\hat{\beta}_{01},\hdots,\hat{\beta}_{0,k-1}$.
\item {\it (Tree Building)}. 
\begin{enumerate}
\item[(a)] For all explanatory variables $x_j,\, j=1,\hdots,p$, fit all the candidate models with one additional split in one of the already built nodes in both components. 
\item[(b)] Select the best model using the $p$-values of the LR test statistics.
\item[(c)] Carry out the permutation test for the selected node (defined by a combination of variable, split point and component) with significance level $\alpha/p$. If significant, fit the selected model and continue with Step~2(a), else continue with Step 3. 
\end{enumerate}
\item {\it (Selected Model)}. Fit the final model with components $\hat{\beta}_{0r}$, $\hat{\betab}$ and $\hat{\gammab}$. 
\end{enumerate}
The final model consists of one or two separate trees, one referring to the location component and one referring to the variance component. In general, the
trees will be different but can also yield the same partitioning. It should be noted that in contrast to the way trees are grown in traditional recursive partitioning  all parameter estimates change if an additional split is performed.  

\section{Further Applications}\label{sec: appl}

\subsubsection*{Biochemists Data}
\begin{figure}[!t]%[!h]
	\centering
		\includegraphics[width=1\textwidth, trim= 0cm 2cm 0cm 0cm]{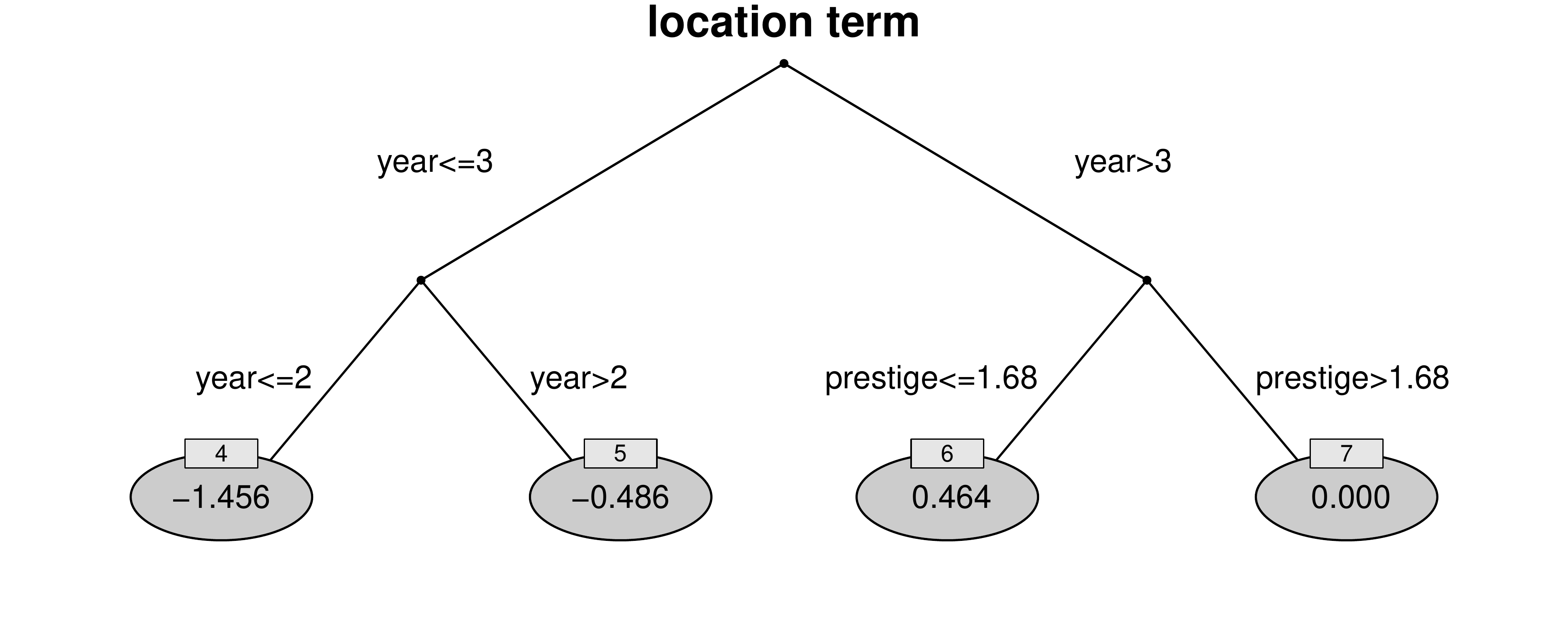}
	\caption{Tree for location term of biochemists example. The parameter estimates~$\hat{\beta}_s$ are given in the terminal nodes.}
  \label{fig:sports}
\end{figure}

\begin{figure}[!t]%[!h]
	\centering
		\includegraphics[width=1\textwidth, trim= 0cm 2cm 0cm 0cm]{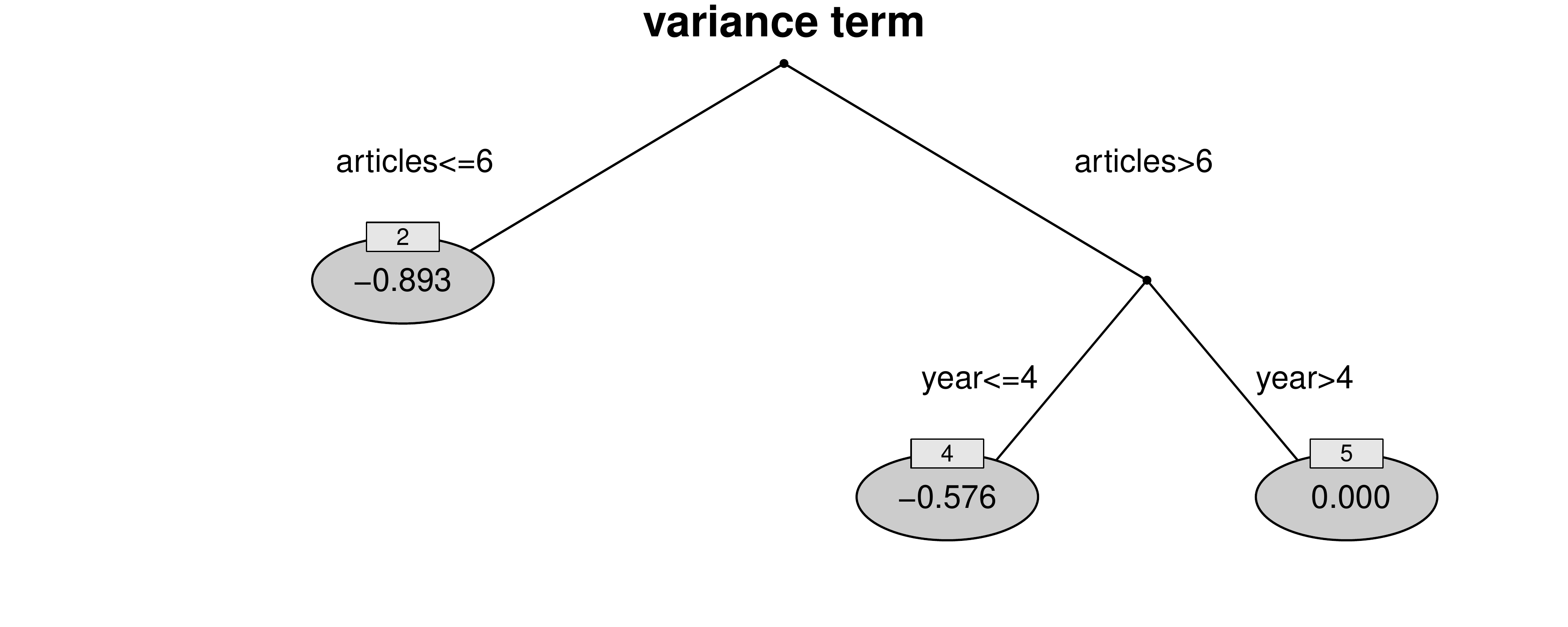}
	\caption{Tree for variance term of biochemists example. The parameter estimates~$\hat{\gamma}_\ell$ are given in the terminal nodes.}
  \label{fig:sports2}
\end{figure}

Let use consider the application used by \citet{allison1999comparing} when investigating the problem if effects of variables differ over gender groups. 
The data set, which has also been used by \citet{long1993rank} and \citet{williams2009using}, investigates
the careers of 301 male and 177 female biochemists (the following description  is adapted from \citealp{allison1999comparing}).
Binary regression is used to predict the probability of promotion to associate
professor from the   assistant professor level (1: no promotion, 2: promotion).
The variables in the model are the number of years since the beginning of the assistant
professorship (years), undergraduate selectivity as a measure of the selectivity of
the colleges where scientists received their bachelor's degrees (select), the number of
articles (articles) representing the cumulative number of articles published by the end of each
person year, and job prestige (prestige)   measuring the prestige of the department in
which scientists were employed. Figures \ref{fig:sports} and \ref{fig:sports2} show the fitted trees for location and variance, respectively. 

While \citet{allison1999comparing} focused on gender as a relevant variable in the variance term, it is seen from the trees that gender does not seem to be very influential; neither in the location term nor in the variance term gender is present. A similar result was obtained by \citet{williams2010fitting}. When he used a stepwise forward  strategy to select variables in the parametric location-scale model the only variable that entered the variance equation was the number of articles. He also made a plausible argument for this by stating that ``there may be little residual
variability among biochemists with few articles (with most of them being denied tenure)
but there may be much more variability among biochemists with more articles (having
many articles may be a necessary but not sufficient condition for tenure).''

It is seen from the trees that the chances of a promotion to associate professor are best for biochemists who have spent at least three years at a department with not the highest prestige (node 6). Applicants with articles $\le 6$ or articles $>6$ in combination with year $\le 4$ seem to form the most homogeneous groups.

\begin{figure}[!t]%[!h]
	\centering
		\includegraphics[width=0.4\textwidth, trim= 2cm 2cm 2cm 0cm]{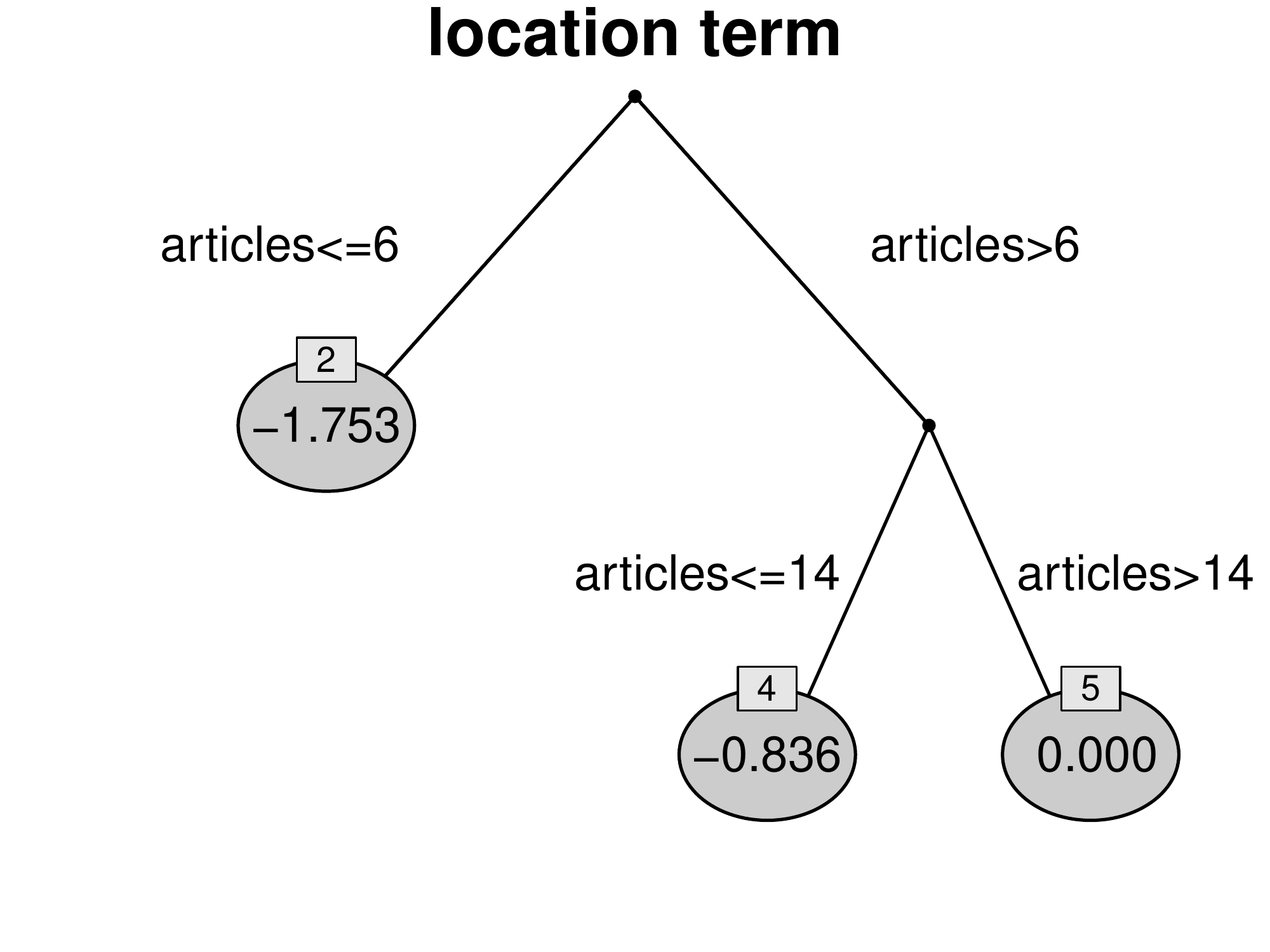}
		\hspace{1.5cm}
    \includegraphics[width=0.4\textwidth, trim= 2cm 2cm 2cm 0cm]{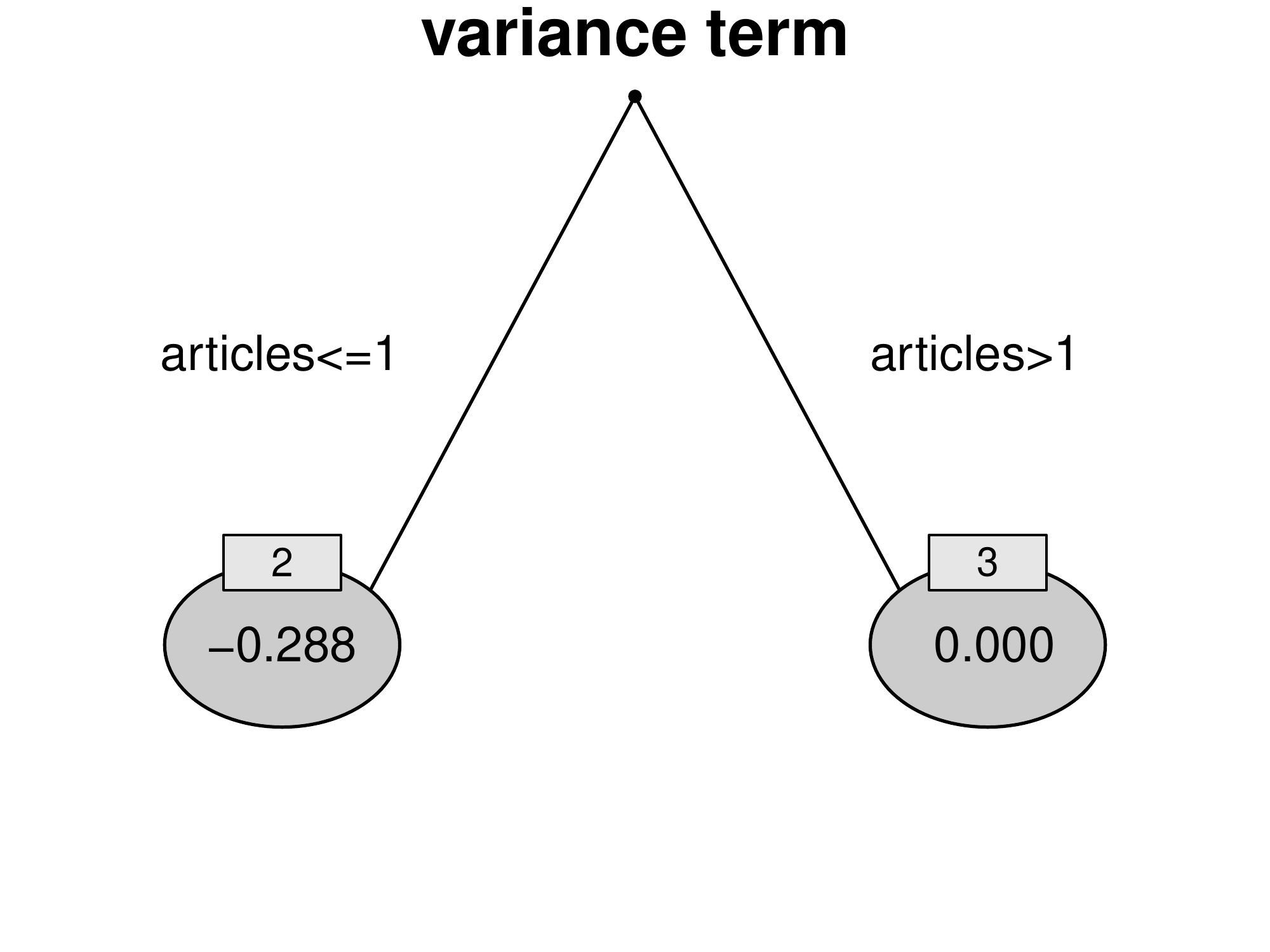}
	\caption{Tree for location (left) and variance (right) of biochemists example with only gender and articles included. The parameter estimates $\hat{\beta}_s$ and $\hat{\gamma}_\ell$ are given in the terminal nodes, respectively.}
  \label{fig:gen}
\end{figure}

To evaluate the issue of unfairness further we fitted trees when only the covariates gender and number of articles are included in the analysis. The corresponding trees are given in Figure  \ref{fig:gen}. It is seen that only the number of articles was found to have an impact on location as well as on variance. There is no indication that gender plays a crucial role for the promotion to associate professor.

\subsubsection*{Retinopathy Data}

In a 6-year followup study on diabetes and retinopathy status
reported by \citet{BenGro:98}
the interesting question was how the retinopathy status is associated
with risk factors. The considered risk factors wer smoking (SM =~1:
smoker, SM = 0: non-smoker),
diabetes duration (DIAB) measured in years, glycosylated hemoglobin
(GH), which is measured in percent, and diastolic blood pressure
(BP) measured in mmHg. The response variable retinopathy status has
three categories (1: no retinopathy; 2: nonproliferative
retinopathy; 3: advanced retinopathy or blind).

\begin{figure}[!t]%[!h]
	\centering
		\includegraphics[width=1\textwidth, trim= 1cm 2cm 1cm 0cm]{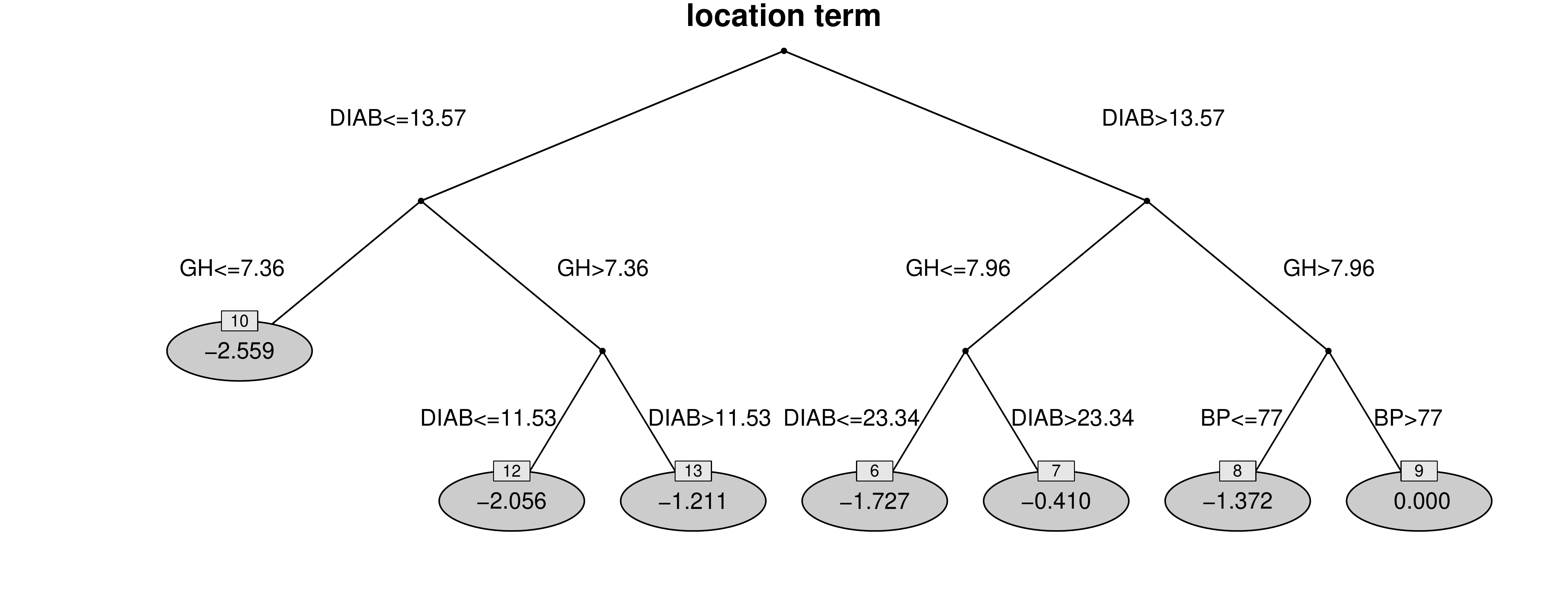}
	\caption{Tree for location term of retinopathy data. The parameter estimates~$\hat{\beta}_s$ are given in the terminal nodes.}
  \label{fig:retfbeta}
\end{figure}

\begin{figure}[!t]%[!h]
	\centering
		\includegraphics[width=0.4\textwidth, trim= 2cm 2cm 2cm 0cm]{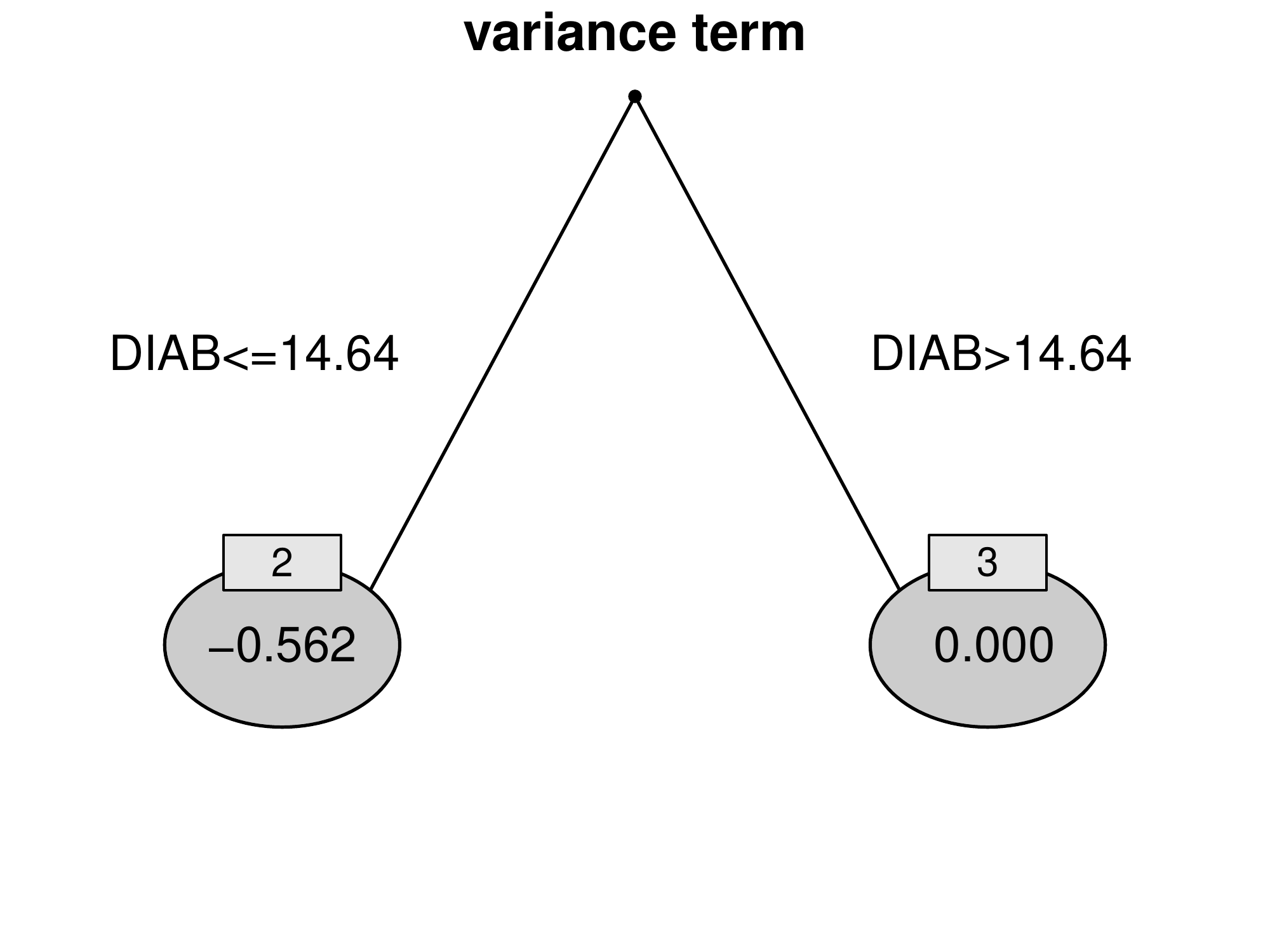}
	\caption{Tree for variance term of retinopathy data. The parameter estimates~$\hat{\gamma}_\ell$ are given in the terminal nodes.}
  \label{fig:retfalpha}
\end{figure}

It is seen from Figure  \ref{fig:retfbeta} that in particular the duration of diabetes is influential followed by glycosylated hemoglobin.
The lowest risk is found in node~10 ($DIAB \le 13.57$, $GH \le 7.36$). Even if $GH > 7.36$ but $DIAB \le 11.53$ the risk is still very low.
The highest risks are found for long duration of diabetes  $DIAB \le 23.34$ in combination with low values of glycosylated hemoglobin $GH \leq 7.96$ (node 7) and in node 9, which combines long diabetes duration and high values of glycosylated hemoglobin and diastolic blood pressure. Figure~\ref{fig:retfalpha} shows that patients with longer duration of diabetes are more homogeneous (sharing higher risk) than patients with lower values of diabetes duration. 


\section{Summary and Concluding Remarks}\label{sec:summ}
Let us summarize the strengths of the proposed tree method.

\begin{itemize}
\item[--]
One obtains two trees, one for the location and one for the variance. Thus, it is clearly seen which variables have an impact on which component. 

\item[--]
The obtained trees have a simple interpretation showing which combinations of variables determine the preference of categories, and which sub populations form more homogeneous or heterogeneous groups. 

\item[--]
By fitting a scale (or variance) component the method avoids misleading effects that may occur if one ignores potential variance heterogeneity. 

\item[--]
As in  all tree-based methods interactions are explicitly  modelled and there is a built-in variable selection procedure.
\end{itemize}
The presented algorithm is constructed such that only variables for which a significant effect can be detected are included. By controlling for the overall significance level the inclusion of irrelevant variables is avoided. It has the effect that the procedure tends to include relatively few variables, in particular if many variables are available. However, the method can also be used in an exploratory way. If one uses a significance level distinctly larger than .05 
one obtains much larger trees, which might hint at further possible interaction effects. Nevertheless, we think it is essential to control for the significance level, which gets lost in many procedures, especially if one first fits trees and then starts pruning as in conventional trees.

An R implementation of the proposed tree-structured model including an auxiliary function to plot the trees, as well as exemplary code to reproduce the illustrative example is available from GitHub~(\texttt{https://github.com/jmober/LocationScaleTree}).

\bibliography{literatur}

\end{document}